\documentclass[9pt,twocolumn,twoside]{osajnl}

\journal{jocn} 

\usepackage{booktabs}

\setboolean{shortarticle}{false}
\title{A broadband vortex beam generation by reflective meta-surface based on metal double-slit resonant ring}

\author[1]{Xufeng Yuan}
\author[1,2]{Chaoying Zhao}

\affil[1]{College of Science,Hangzhou Dianzi University, Zhejiang 310018, China.}
\affil[2]{State Key Laboratory of Quantum Optics and Quantum Optics Devices, Institute of Opto-Electronics,Shanxi University, Taiyuan 030006, China}

\affil[*]{email@my-email.com}

\begin{abstract}
 Recently, meta-surface(MS) has emerged as a promising alternative method for generating vortex waves. At the same time, MS also face the problem of narrow bandwidth, in order to obtain a board bandwidth, the MS unit cells structure become more and more complex, which will deduce many inconveniences to the preparation process of MS device. Therefore, we want to design a simple MS unit cell with a multi-frequency selection. In this paper, based on the principle of geometric phase, we design a simple reflective MS unit cell based on metal double-slit resonant ring. We elaborate on the resonance mechanism of the MS unit cell. Under the normal incidence of circularly polarized($CP$) waves, the reflection coefficient of the same polarization was greater than $85\%$. By rotating the orientation angle of the resonator on the MS unit cell, the continuous $2\pi$ phase coverage was satisfied in the frequency range of $0.52THz-1.1THz$, and the relative bandwidth becomes $71.6\%$. Based on this, we construct a vortex generator by using a $15X15$ MS unit array. The right-handed circularly polarized waves ($RCP$) and left-handed circularly polarized waves ($LCP$) are separately incident on MS with topological charges of $l=+1,+2,+3$ under multiple resonant frequencies. The generated $RCP$ vortex wave with topological charges of $l=-1,-2,-3$ and the generated $LCP$ vortex wave with topological charges of $l=+1,+2,+3$. The numerical simulation results exhibit our designed MS with multiple resonance outcomes can achieve a multi-broadband operation and generate a wide-band vortex beam. In addition, we also calculate the pattern purity. Through theoretical analysis and numerical simulation, we prove that our designed MS can generate a broadband vortex wave.
 
\end{abstract}

\setboolean{displaycopyright}{false} % Do not include copyright or licensing information in submission.

\begin{document}

\maketitle

\section{Introduction}
Optical vortices were first introduced in 1989\cite{46}, and in 1992, Allen et. al.\cite{20} conducted an experiment on the paraxial propagation of vortex waves. This experiment has confirmed that the vortex wave can carried an orbital angular momentum (OAM) of $l\hbar$, where $\hbar$ is the reduced Planck constant, and $l$ is the topological charge of the vortex wave. The experiment established the existence and fundamental properties of optical vortices, revealing that vortex waves with different orders exhibit orthogonality, preventing interference among them. This unique property renders vortex waves have a highly valuable in high-capacity optical communication systems.
Utilizing the orthogonality of vortex waves, it can be used as an efficient information carrier in modular division multiplexing (MDM) systems\cite{47}, MDM integrates polarization division multiplexing (PDM)\cite{47} and wavelength division multiplexing (WDM) technologies\cite{31} to significantly boost channel capacity. There are some applications in coding\cite{52} and quantum entanglement\cite{53}, quantum communication\cite{54}, as well as micro-nano manipulation\cite{55}.

The generation of vortex waves holds significance for the exploration of their characteristics and potential applications. Traditionally, vortex wave generation relies on optical components such as Spiral phase plates (SPPs)\cite{27}, spatial wave modulators (SLMs)\cite{26}, fork gratings \cite{25}, and array antennas\cite{45}. Nevertheless, these conventional optical elements and devices encounter challenges such as complexity, reduced efficiency, and limited wavelength operation, thereby impeding the advancement of vortex wave research and utilization.

In recent years, meta-surface (MS) have emerged as a promising alternative method for generating vortex waves. MS are two-dimensional meta-materials composed of periodic sub-wavelength structures that can be customized to provide integrated, broadband, efficient, and versatile devices. Until now, the applications of MS in the field of optics encompass polarization conversion\cite{49}, holography\cite{28}, focusing\cite{29}, complete absorption\cite{30}, wave shaping\cite{48}, and various other functions.
Furthermore, terahertz waves have found widespread applications in biology, medicine\cite{37,38,39}, and communication\cite{6,7}, among other domains, due to their abundant spectrum resources, non-destructive nature, and high resolution. The integration of MS with terahertz waves exhibits significant promise in the generation of high-frequency, efficient vortex waves.

In 2011, Yu et. al. \cite{2}derived a generalized Snell's law for MS by leveraging Fermat's principle. They designed an MS consisting of eight $45^{\circ}$ V-shaped structural units, aimed at generating a vortex wave carrying a topological charge of $l$=1 in the near-infrared spectrum. Nevertheless, the designed MS exhibited low efficiency.
In 2020, Li \cite{41} and Zhang proposed a MS generator for OAM based on geometric phase \cite{43}. This MS was capable of generating vortex waves with topological charges of $l=\pm 1$ and $l=\pm 2$ across a frequency range of $0.3THz$ to $0.45THz$. Furthermore, recent advancements have extended the capabilities of vortex generators, bridging the gap from single-frequency to multi-frequency and wider-band applications\cite{14,15,16,17}. In 2021, Guo et. al. designed a  geometric -phase-based reflective MS for generating vortex wave over a broadband range of $0.9THz$ to $1.8THz$ \cite{40} .
In this work, we want to  design a geometric phase reflective MS, capable of generating vortex wave within the frequency range of $0.52THz$ to $1.1THz$. The MS unit cell comprises a metal-insulator-metal (MIM) configuration, which incorporates a top metal double-slit resonant ring (DSRR), a middle dielectric layer, and a bottom metal substrate. 

When circularly polarized (CP) wave vertically incident on the unit cell, continuous phase modulation of the reflected CP wave can be achieved within the target frequency range, covering a phase range of 0 to $2\pi$. Additionally, the vortex generator formed by specially arranged unit cells generates circularly polarized vortex waves with topological charges of $l=±1,±2,±3$ within the broadband of interest. We obtain the reflection phase and amplitude of the vortex wave under various topological charges through numerical simulations and calculate the mode purity. 
The numerical results align well with the anticipated theoretical outcomes, demonstrating that the proposed MS can generate vortex waves of superior quality with distinct topological charges across the target broadband. The simplicity of the proposed structure offers great potential for the integration of multi-functional MS, which can be easily extended to other frequency ranges.

\section{The design and analysis of Meta-surface unit cells}
\subsection{Design and theoretical analysis}
Fig. 1(a) depicts the schematic diagram of the proposed MS wide-band vortex generator. Upon irradiation of CP wave onto the MS, the reflected circular polarization component with the same polarization state as the incident wave is converted into vortex wave.
Figs.1(b) and 1(c) illustrate the top and side views of the MS unit cell structure, respectively, which employs a MIM configuration. The middle medium layer consists of FR-4 material, exhibiting a lossy characteristic with a permittivity of 4.3 and a loss tangent of 0.025. The top DSRR and the bottom substrate are crafted from copper, exhibiting a conductivity of $5.8\times10^7 S/m$. The optimized geometric parameters of the designed structure are as follows: $p=84\mu m$, $a=b=10\mu m$, $d=25\mu m$, $h_{1}=0.5\mu m$, $h_{2}=36\mu m$, $\theta$ represents the rotation angle between the DSRR and the $x$-axis.

\begin{figure}[htbp]
\centering{
\includegraphics[width=\linewidth]{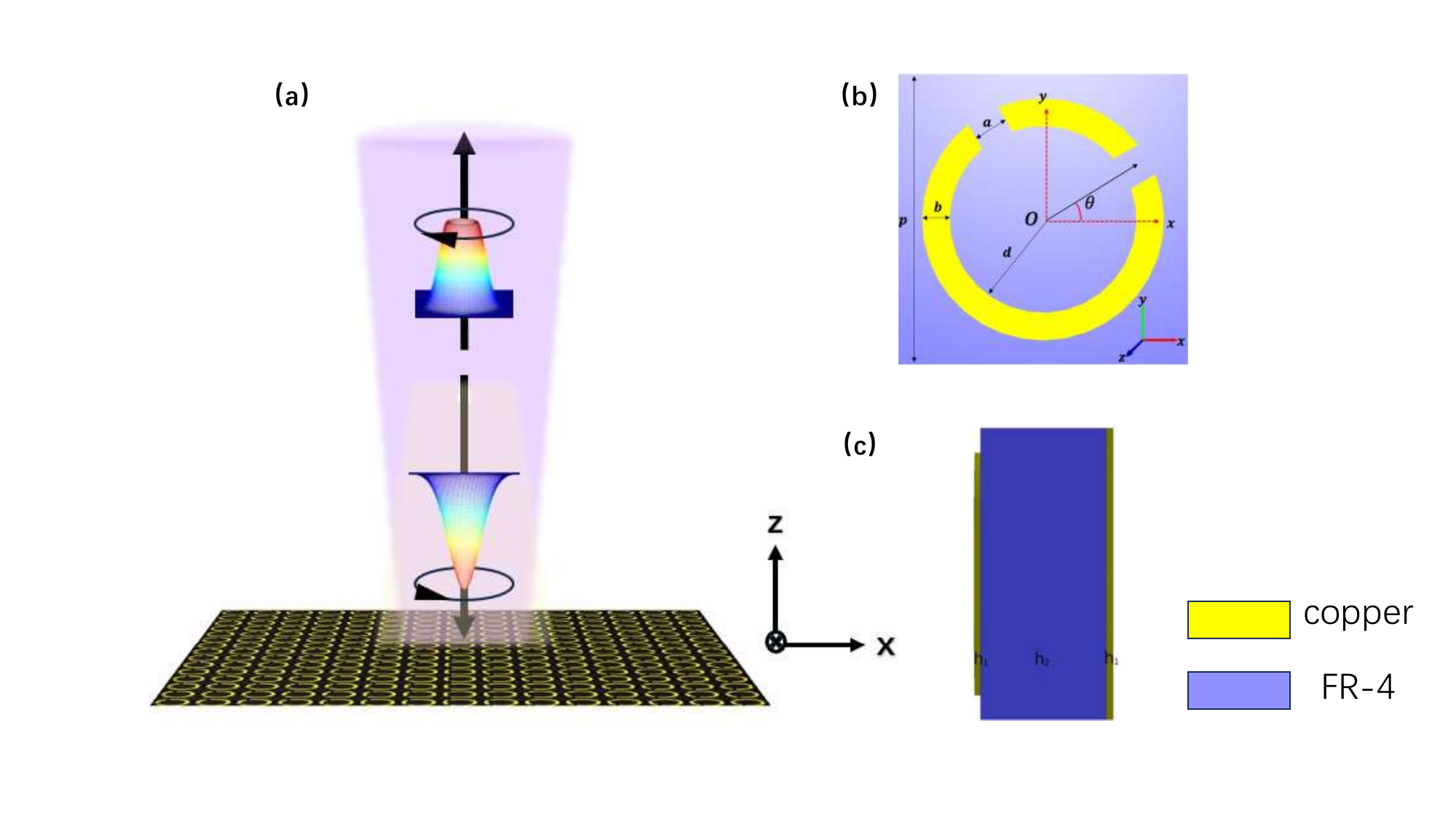}
\label{fig:FIG.1.a}}
\caption{(a)Schematic diagram of the reflected MS: The reflected $CP$ wave carries OAM is consistent with the polarization state of the incident $CP$ wave; (b)Top view of unit cell; (c)Side view of unit cell}
\end{figure}

Using the principle of geometric phase, a continuous  $2\theta$, phase shift is achieved by adjusting the orientation angle $\theta$, of the DSRR. When the MS is illuminated by $CP$ waves, this process can be derived from the linear polarization ($LP$) reflection coefficient matrix through the rotation matrix  R($\theta$) and the transformation matrix $\Lambda$,as detailed in Eqs.(1) to (3).

\begin{align}
\begin{split}
J_{\theta}^{LP}&=R(\theta)^{-1}JR(\theta)\\&=
\begin{pmatrix}cos\theta&sin\theta \\
  -sin\theta &cos\theta
\end{pmatrix}^{-1} \begin{pmatrix}
 r_{xx}  &  r_{xy}  \\
r_{yx}  & r_{yy}  
\end{pmatrix} \begin{pmatrix}cos\theta&sin\theta \\
  -sin\theta&cos\theta
\end{pmatrix}
\label{eq:1}
\end{split}
\end{align}
The Jones matrix $J$ are used to describe the modulation effect of unit cell devices on linear polarized waves, which also serves as the LP reflection coefficient matrix for the MS cell. $R(\theta)$ represents the rotation matrix. As detailed in Eq.(1), by adjusting the orientation angle of DSRR, a new transmission matrix $J_{\theta}^{LP}$ is derived. The superscript $LP$ signifies that this transmission matrix is tailored for linear polarization incidence. Given that the MS unit introduced in this paper can convert linear polarization to cross polarization, the most ideal situation is that: $r_{xx}=r_{yy}=0$, $r_{xy}=r_{yx}=1$ .

In the case of circular polarization incidence, the device transmission matrix $J_{\theta}^{CP}$ can be obtained by the transformation matrix $\Lambda =\frac{1}{\sqrt{2} } \begin{bmatrix}
 1 & 1\\
 i &-i
\end{bmatrix}$ , 
\begin{align}
\begin{split}
J_{\theta}^{CP}=\Lambda ^{-1} J_{\theta}^{LP}\Lambda =\frac{1}{2} \begin{bmatrix}
 i(r_{xy}-r_{yx})  & -i(r_{xy}+r_{yx})e^{-i2\theta }\\ 
i(r_{xy}+r_{yx})e^{i2\theta } &  i(r_{yx} -r_{xy} )
\end{bmatrix}
\end{split}
\end{align}

The superscript $CP$ is used to designate that the transmission matrix is appropriate for $CP$ incidence.

When using $CP$ wave incident:
\begin{align}
\begin{split}
\begin{bmatrix}
 E_{L}\\E_{R}
\end{bmatrix}=\frac{1}{2} \begin{bmatrix}
 i(r_{xy}-r_{yx})  & -i(r_{xy}+r_{yx})e^{-i2\theta }\\ 
i(r_{xy}+r_{yx})e^{i2\theta } &  i(r_{yx} -r_{xy} )
\end{bmatrix}\begin{bmatrix}
 E_{R}\\E_{L}
\end{bmatrix}
\end{split}
\end{align}
The definition of reflected $LCP$ and reflected $RCP$ is:
\begin{align}
\begin{split}
\left \{
\begin{array}{lr}
E_{L}=(E_{x} -iE_{y})/\sqrt{2} \\
E_{R}=(E_{x} +iE_{y})/\sqrt{2} 
\end{array}
\right.
\end{split}
\end{align}

From Eq.(3), it is evident that MS generates an co-polarized $CP$ component that exhibits an added phase shift of ($\pm2\theta$). Consequently, theoretical analysis confirms that DSRR attains to $0-2\theta$ phase coverage through the rotation of the orientation Angle from $0-\theta$. When left-circularly polarized waves and right-circularly polarized waves are incident, respectively, the additional phase carried by the co-polarization term in the reflected wave maintains the same magnitude but has opposing signs. This underpins the capability of our designed MS to generate topological vortex waves with opposing symbols.

\subsection{Numerical simulation and resonance mechanism analysis of unit cell}

CST MICROWAVE STUDIO was utilized for the numerical simulation of the unit cell, and a $LP$ wave incident was set.  Upon examination of Fig.2, it is evident that the maximum values of the co-polarization components, $r_{xx}$ and $r_{yy}$, are less than 0.37, while the cross-polarization components, $r_{xy}$ and $r_{yx}$, exceed 0.85 within the frequency range of $0.52THz-1.1THz$, $r_{xy}$ represents the reflection coefficient when $y-LP$ wave is incident and the outgoing wave is $x-LP$ wave. The definition of $r_{yx}$, $r_{xx}$, $r_{yy}$ in a matrix is similar to $r_{xy}$. Polarization conversion rate (PCR=$\frac{\left|co-polarization component\right|^2}{\left|co-polarization component\right|^2+\left|cross-polarization components\right|^2}$) was used to evaluate the polarization conversion level of MS unit cell. Fig.2 demonstrates that within the relevant frequency range, the PCR attains to a minimum of approximately $85\%$, while the polarization conversion efficiency hit $100\%$ at three distinct points. 

\begin{figure}[htbp]
\centering{
\includegraphics[width=\linewidth]{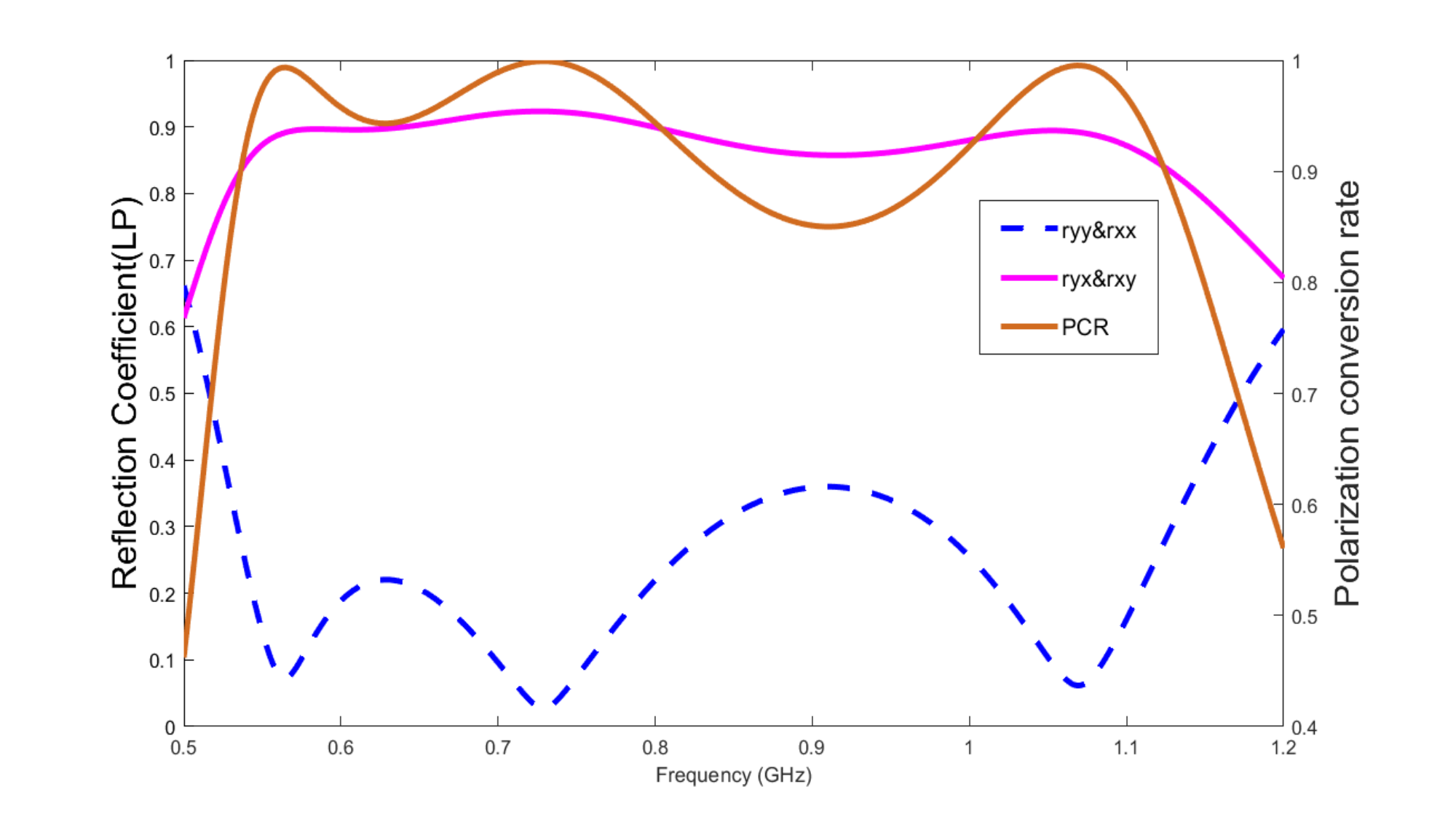}
\label{fig:FIG.2.}}
\caption{Reflection coefficient and PCR when $LP$ wave incident vertically on unit cell.}
\end{figure}
The effective electromagnetic parameters of the MS are derived using the S-parameter inversion method. As depicted in Fig.3, the effective permeability (\(\mu_{eff}\)) of the structure closely resembles the effective permittivity (\(\varepsilon_{eff}\)) across the frequency range of $0.52THz$ to $1.1THz$. By considering impedance matching conditions, where \(z_{eff}=\sqrt{\mu_{0}\mu_{eff}/{\varepsilon_{0}\varepsilon_{eff}}}\) equals to the spatial wave impedance \(z_{0}\), with \(\mu_{0}\) and \(\varepsilon_{0}\) representing vacuum permeability and vacuum permittivity, respectively. It can be observed that, within this frequency range, the equivalent wave impedance \(z_{eff}\) aligns closely with the spatial wave impedance \(z_{0}\) at frequencies of $0.56THz$, $0.73THz$, and $1.06THz$. At these frequencies, the reflection of co-polarized waves is minimized, resulting in a maximum PCR. Conversely, outside the relevant frequency range, the disparity between effective permeability and effective permittivity increases, leading to enhanced reflection of co-polarized waves and a reduction in PCR. This observation suggests that the MS effectively converts incident linearly polarized waves into their corresponding cross-polarized waves.

\begin{figure}[htbp]
\centering{
\includegraphics[width=\linewidth]{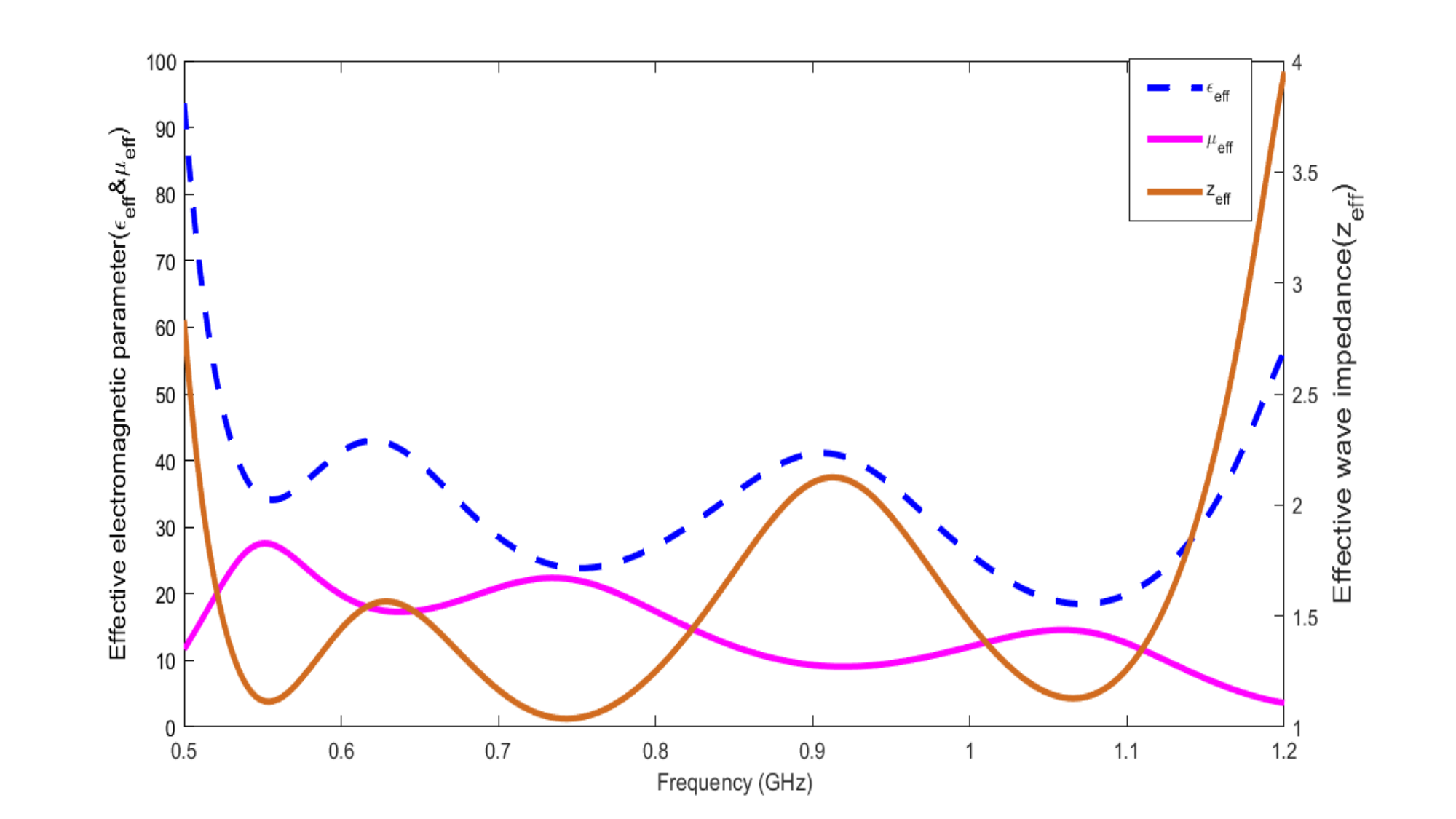}
\label{fig:FIG.3.}}
\caption{Effective permeability, effective dielectric constant and effective wave impedance}
\end{figure}

To clarify the resonance mechanism of MS unit cell ,using the example of an $LCP$ wave incident vertically on a DSRR ($\theta$=0), we analyze how the anisotropy structure of the top layer of the MS and the overall MIM structure impact the surface current distribution under the irradiation of the incident electromagnetic wave. Specifically, Fig.4 presents the instantaneous surface current and electric field at three resonant frequencies: $0.56THz$, $0.73THz$, and $1.07THz$.

\begin{figure}[ht]
\centering{
\includegraphics[width=\linewidth]{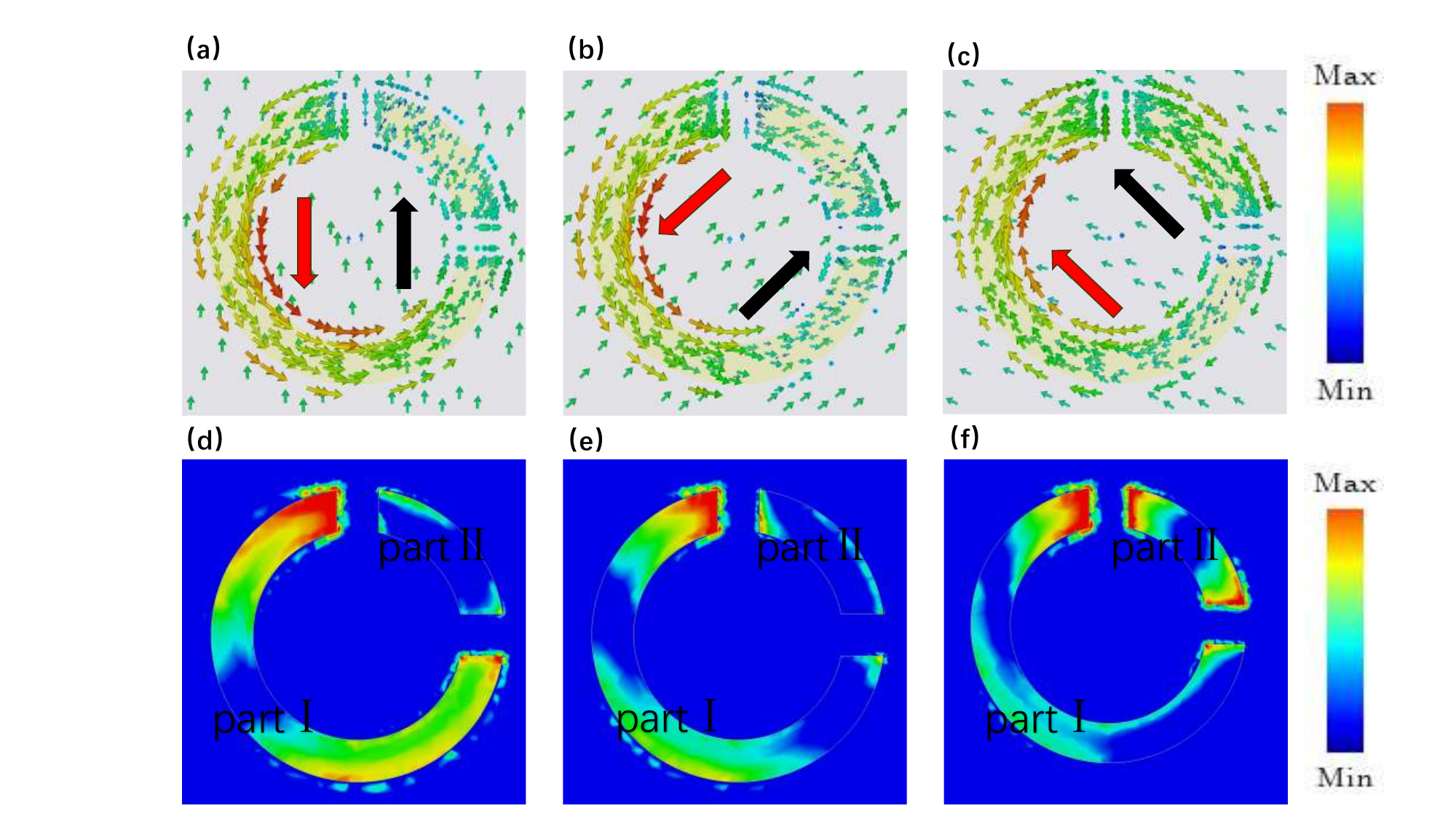}}
\caption {Instantaneous surface current distribution:(a),(b),(c) and electric field:(d),(e),(f) of MS unit cell at: (a),(d) $0.56THz$ and (b),(e) $0.73THz$ (c),(f) $1.07THz$.}
\end{figure}

In Figs.4(a) and 4(b), the surface current distribution of the top metal ring and the bottom metal plate is in opposite directions, thus forming a complete current loop. The above process will gives rise to magnetic dipole characteristics (such as magnetic resonance) within the MS unit. In contrast, Fig. 4(c) depicts a situation where the surface current direction of both the top and bottom elements is identical, leading to the emergence of electric dipole characteristics within the MS unit and subsequently electrical resonance. When a $RCP$ wave is irradiated vertically, the direction of the instantaneous surface current at each frequency is diametrically opposed to that of the $LCP$ wave.
Figs. 4(d), 4(e), and 4(f) depicts the electric field distribution of the unit cell at three resonant frequencies, respectively, and revealing that the resonant positions of the unit cell vary across frequencies. At the lower frequencies of $0.56THz$ and $0.73THz$, part \uppercase\expandafter{\romannumeral1} hardly participates in resonance, whereas part \uppercase\expandafter{\romannumeral1}  and part \uppercase\expandafter{\romannumeral2} resonate at $1.07THz$. As the incident wave induces charge accumulation at the opening, the charge motion generates a current. Think of DSRR as a simple LC oscillator circuit,the ring serves as the inductance L, the opening as the capacitance C, using the approximate resonant frequency equation $f\sim \frac{1}{2\pi \sqrt{LC}}$, combined with the surface current distribution lengths shown in Figs. 4(a), 4(b), and 4(c), the equivalent inductance is proportional to the current length, the equivalent inductance is inversely proportional to the frequency, this leads to the generation of different surface currents and electric fields at different frequencies.The simulation results align with the theoretical.

We employed the CST MICROWAVE STUDIO software to conduct numerical simulations,to verify that the designed MS unit cells can achieve a phase jump of twice the orientation angle size in the reflected co-polarized component by rotating the DSRR orientation angle The boundary conditions were set as follows: $Z_{min}$ was set as electric ($E_t=0$), $Z_{max}$ was set as open (with added space), and both the $x$-axis and $y$-axis were set as the unit cell. We utilized circularly polarized waves ($LCP$ and $RCP$) incident normally on the MS unit cells.

\begin{figure}[ht]
\centering{
\includegraphics[width=\linewidth,height=.6\linewidth]{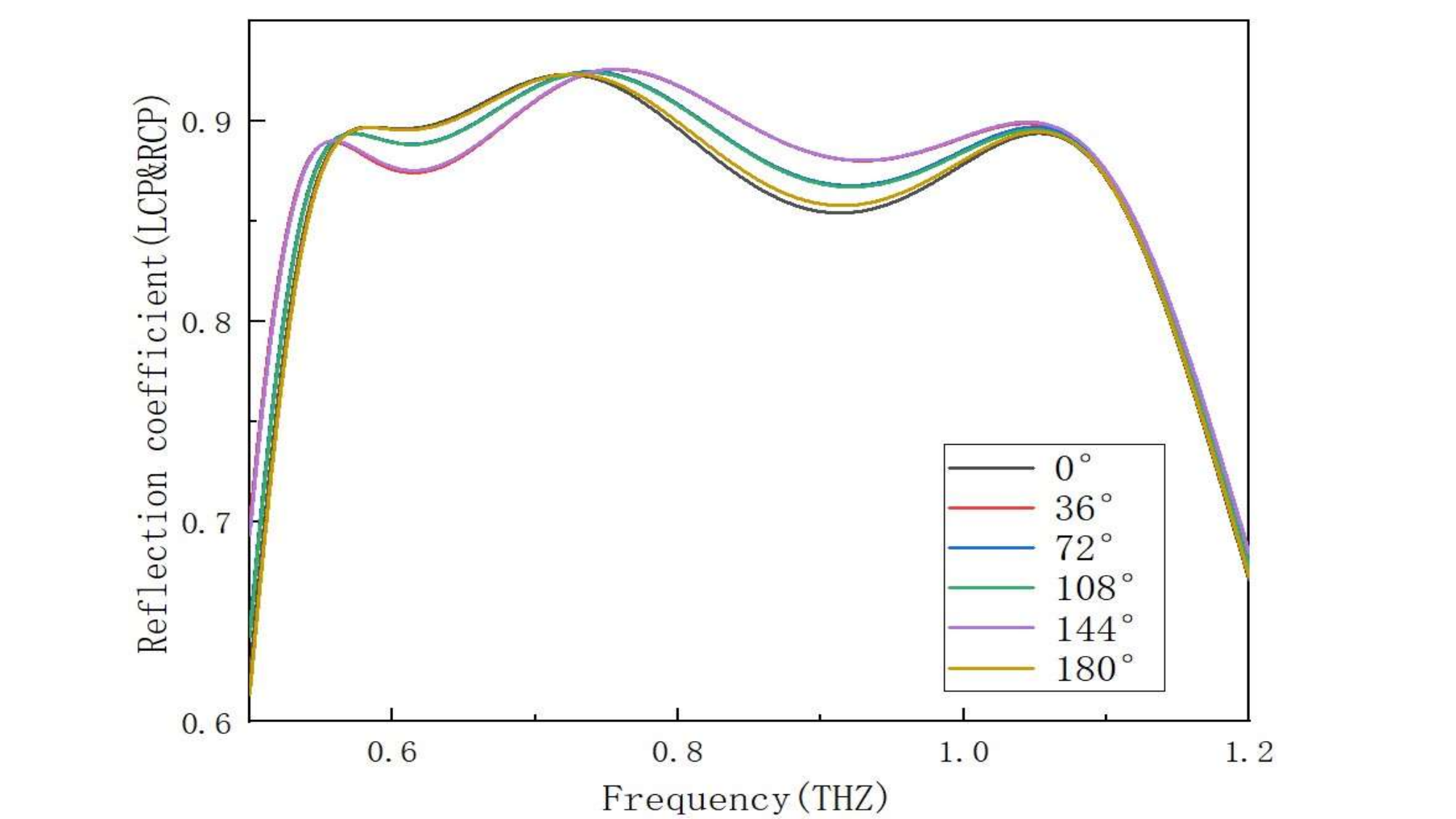}}
\label{fig:FIG.5.}
\caption{Reflection coefficient when  $LCP$ and $RCP$ normally incident on unit cells with different orientation angle.}
\end{figure}

\begin{figure}[ht]
\centering{
\includegraphics[width=\linewidth,height=.6\linewidth]{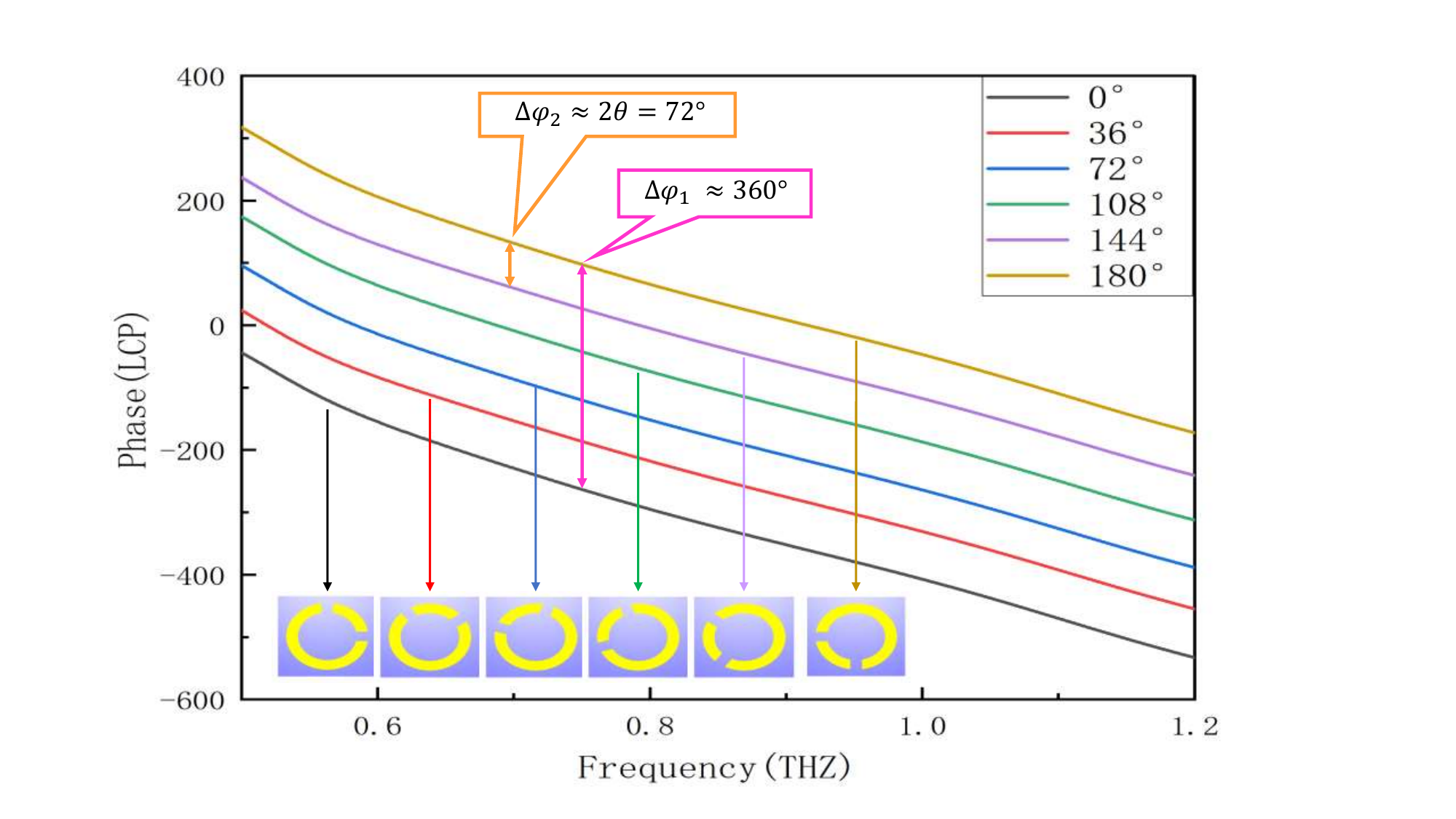}\put(-220,120){(a)}
\label{fig:FIG.6.a}
\includegraphics[width=\linewidth,height=.6\linewidth]{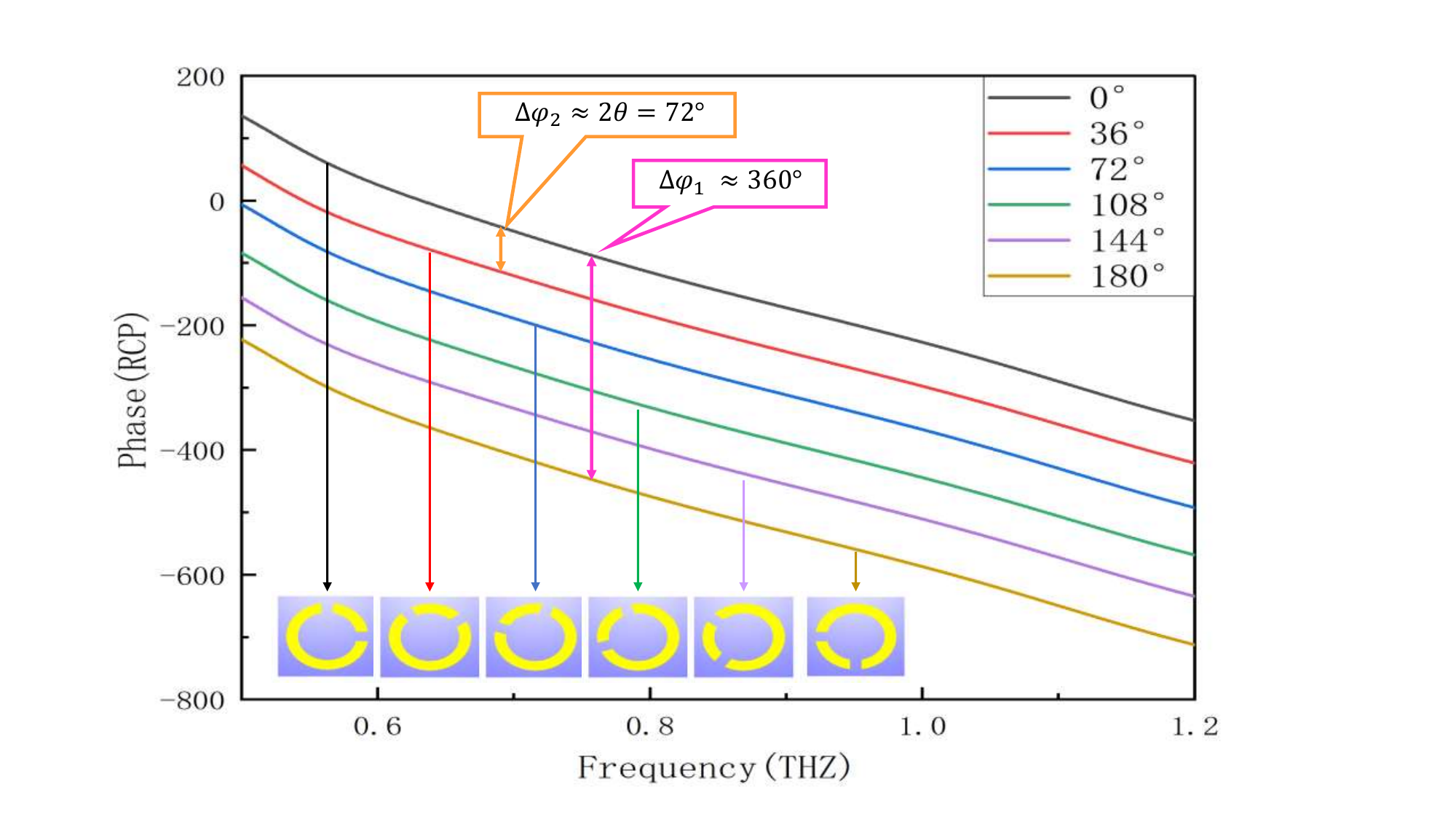}\put(-220,120){(b)}}
\label{fig:FIG.6.b}
\caption{Phase of reflected wave when (a)$LCP$ and (b)$RCP$ normally incident on unit cells with different orientation angle.}
\end{figure}

When the orientation angle $\theta$ of DSRR varies from $0^{\circ}$ to $180^{\circ}$ in steps of $36^{\circ}$, Fig.5 illustrates the amplitude of the reflected co-polarized CP wave within the frequency band of interest. Notably, when both $LCP$ and $RCP$ waves incident under identical orientation conditions, the amplitude of the reflected co-polarized component remains the same. Additionally, Fig.6 depicts the phase of the reflected co-polarized $CP$ wave. 
Reflection phase $\Delta \varphi _{2} =2\theta=72^{\circ}$ corresponding to a $36^{\circ}$ change in DSRR orientation angle $\theta$, which is consistent with Eq.(3),
$\Delta \varphi _{1} $ represents that as the DSRR orientation angle $\theta$ varies from $0^{\circ}$ to $180^{\circ}$ with an interval of $36^{\circ}$, the phase of the reflected co-polarized wave varies from $0^{\circ}$ to $360^{\circ}$. In the frequency range of interest, a phase coverage of 0-2$\pi$ is achieved and remains relatively stable.
This suggests that vortex generators with $l=1$ or higher can be realized through careful arrangement of MS units. Numerical simulations confirm that, within the frequency range of $0.52THz-1.1THz$, the reflection amplitude remains above $85\%$, relative bandwidth of $71.6\%$.

In the theoretical analysis and simulation present above, we have demonstrated that a co-polarized CP wave can effectively achieve carrying a phase of $2\theta$ by rotating DSRR orientation angle $\theta$ across a broad frequency range of $0.52THz-1.1THz$. Furthermore, the high reflection coefficient obtained indicates the excellent structural performance of our MS unit cell.

\section{Simulation and Analysis of Reflective Meta-surfaces for Generating Vortex Waves }
The reflective MS designed by us is comprised of 15x15 unit cells, MS measuring 1260$\mu m$ x 1260$\mu m$.
We simulate the generation of the $l$th-order vortex wave by a reflective MS. In order to achieve the phase distribution of vortex beam $e^{il\psi}$ on the cross-section, the corresponding relationship between the phase distribution $\psi_{l}$ required at each coordinate $(x,y)$ is as shown in Eq.(5), the phase $\psi_{l}$ required for each position , which is obtained by rotating the DSRR orientation angle $\theta=\psi_{l}/2$.

\begin{align}
\begin{split}
\psi _{l}=arctan(\frac{y}{x} )\times l
\end{split}
\end{align}

The phase distribution of a reflective MS with topological charges of $l=+1,+2,+3$ is presented in Fig. 7(a), 7(b), and 7(c), respectively. 
In addition, Fig. 7(d) shows a reflective MS with $l=+1$ and indicates the rotation of DSRR at each position on the MS. Each position of DSRR has a different orientation angle to meet the phase distribution in Fig.7(a).
Referring to Fig. 6, $LCP$ and $RCP$ waves incident on a cell rotating continuously from $0^{\circ}$ to $180^{\circ}$ introduce phases of $+360^{\circ}$ and $-360^{\circ}$, respectively. When $LCP$ and $RCP$ waves interact with the reflective MS of $l=+1$, the reflecting components of $LCP$ and $RCP$ carry topological charges of $l=+1$ and $l=-1$, respectively.
The preparation of reflective MS can be achieved through mask lithography \cite{19}. This machining method is not only simple and repeatable but also ensures that the metal structure maintains its shape with minimal thickness. 

\begin{figure}[ht]
\centering{
\includegraphics[width=\linewidth]{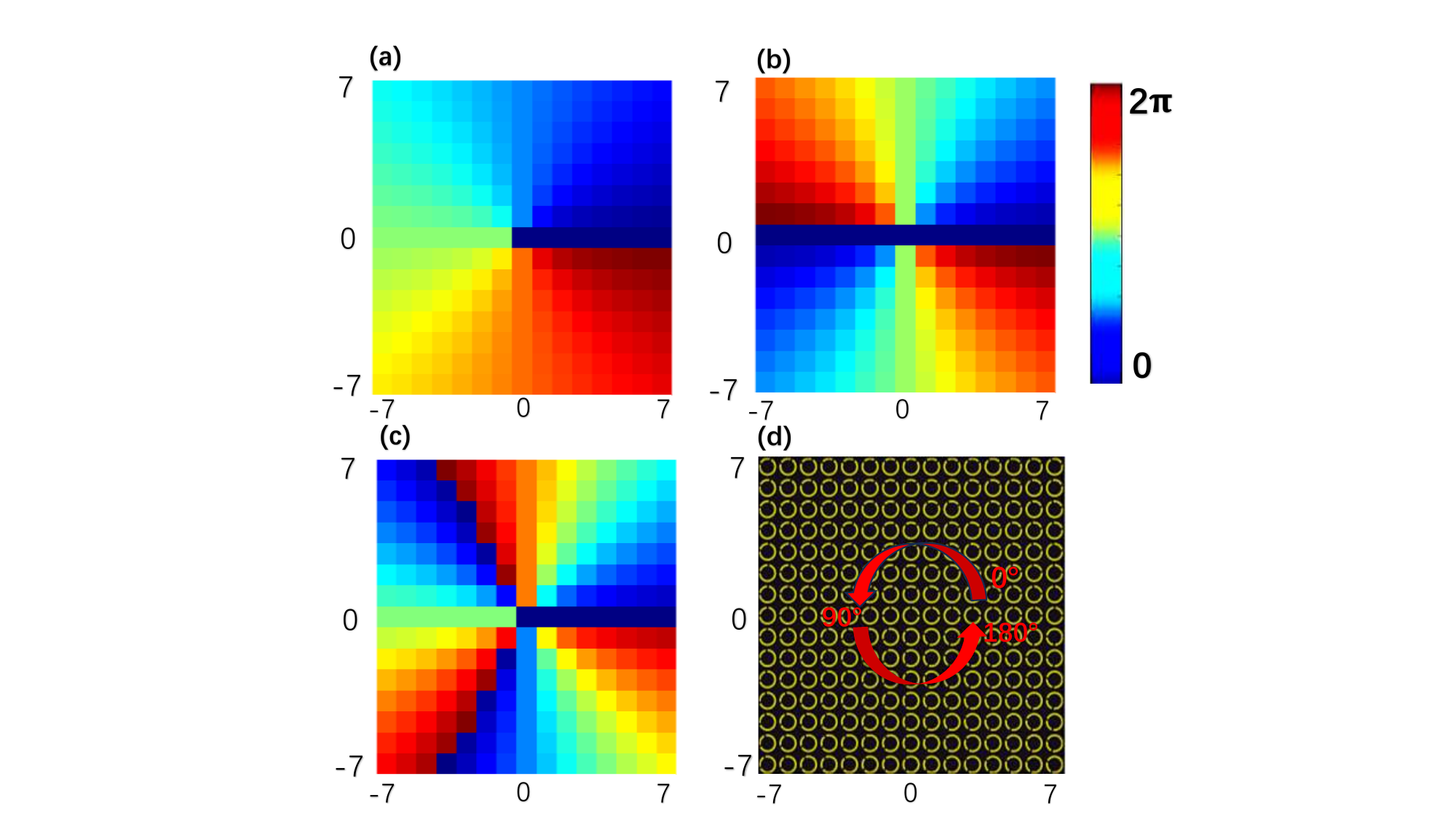}
\label{fig:FIG.7.a}}
\caption{MS phase distribution:(a)$l=+1$, (b)$l=+2$, (c)$l=+3$; Reflective MS with a topological charge of $+1$.}
\end{figure}

We have selected the Gaussian wave ($LCP$ and $RCP$), as the excitation source. This source is positioned at a distance of $600\mu m$ from the MS. Additionally, we have established the background conditions and chosen three frequencies: $0.7THz$, $0.9THz$, and $1.1THz$, for conducting full-wave simulations.

 When $RCP$ Gaussian wave is incident vertically on the $l=+1$ order MS, the amplitude and phase of the reflected wave are measured at different positions(L) and frequencies($f$).
\begin{figure}[ht]
\centering{
\includegraphics[width=\linewidth,]{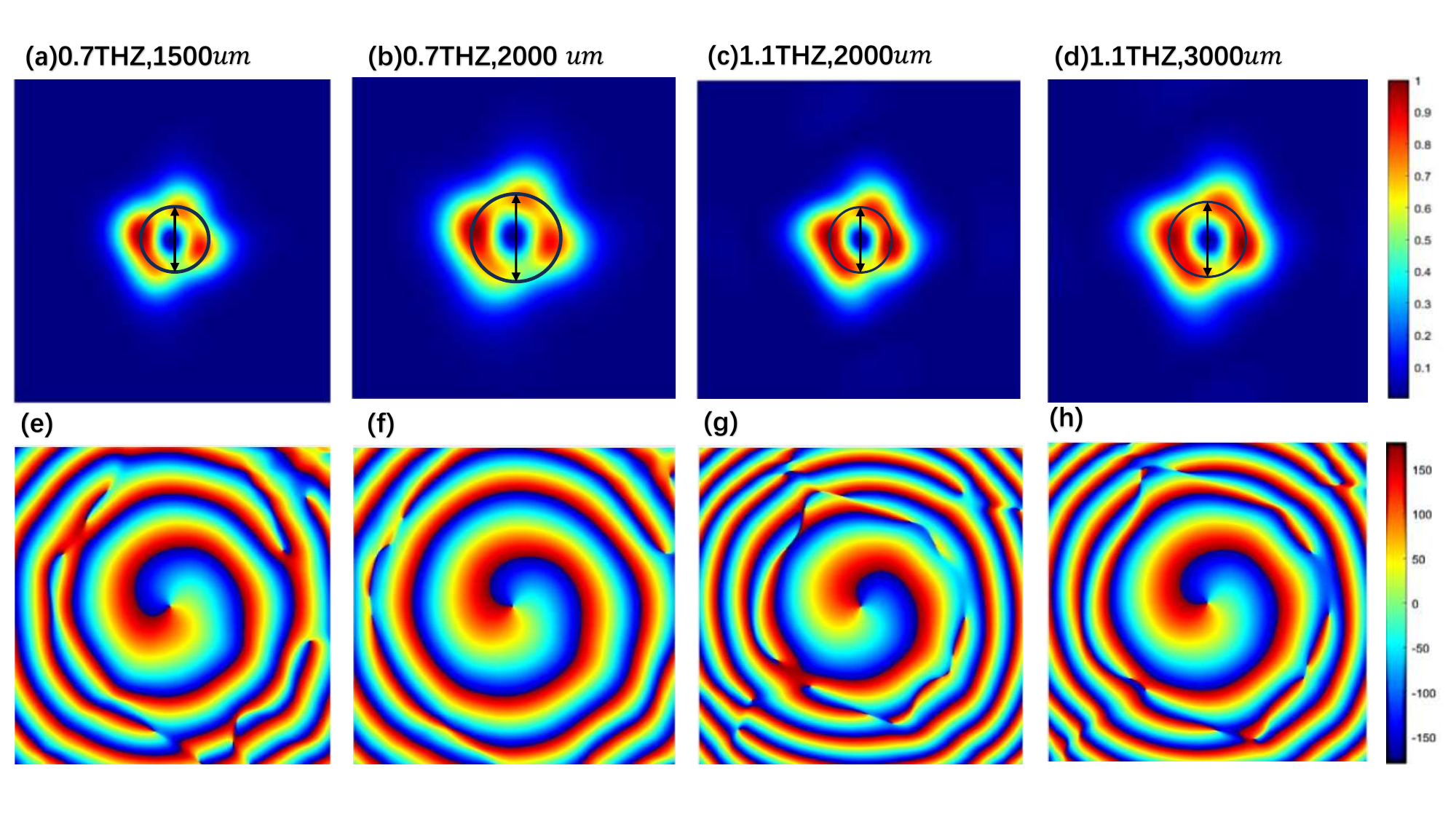}}
\caption{ The normalized reflection amplitudes:(a) $L=1500\mu m$ at $f=0.7THz$,(b) $L=2000\mu m$ at $f=0.7THz$, (c) $L=2000\mu m$ at $f=1.1THz$,(d)$L=3000\mu m$ at $f=1.1THz$; Reflection phase: (e), (f), (g), and (h) correspond to (a), (b), (c), and (d), respectively.}
\label{fig:FIG.8.}
\end{figure}

Fig. 8 demonstrates the successful conversion of the $RCP$ reflected by the MS into a vortex wave carrying a topological charge of $l=-1$, which aligns with our expectations.The reflection amplitude's central intensity  is zero due to the phase singularity, resulting in a doughnut-like shape. Additionally, the corresponding reflection phase exhibits a spiral arm pattern. When comparing Figs. 8(a), 8(b), 8(c), and 8(d), it becomes evident that the OAM waves broaden as the transmission distance increases at a given frequency,
In order to better understand this concept, we draw a circle with the distance from the center to the amplitude peak as the radius.  When comparing Figs.8(b) and 8(c), it becomes evident that, at the same transmission distance, the OAM wave undergoes a significant widening as the wavelength increases.

 Subsequently, we simulate a reflective MS with a value of $l=+2$, and the corresponding results are depicted in Fig.9.
\begin{figure}[ht]
\centering{
\includegraphics[width=\linewidth]{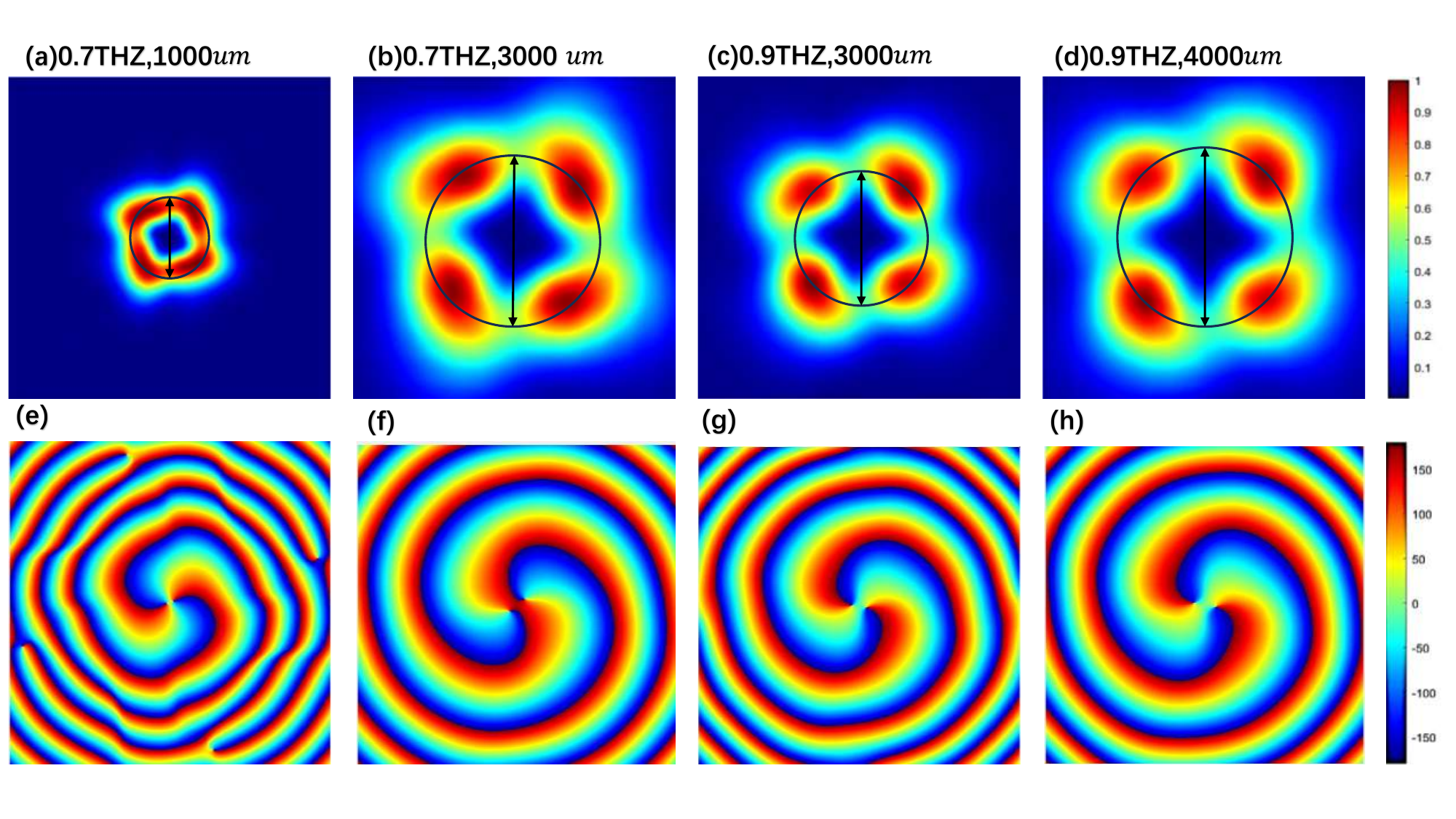}}
\caption{The normalized reflection amplitudes are presented for the following conditions: (a)$l=-2$ with $L=1000\mu m$ at $f=0.7THz$, (b)$l=-2$ with $L=3000\mu m$ at $f=0.7THz$, (c)$l=-2$ with $L=3000\mu m$ at $f=0.9THz$, and (d)$l=-2$ with $L=4000\mu m$ at $f=0.9THz$. The corresponding reflection phases for (a), (b), (c), and (d) are shown in (e), (f), (g), and (h), respectively.}
\label{fig:FIG.9.}
\end{figure}

As anticipated, the incident $RCP$ transforms into a $RCP$ vortex wave carrying $l=-2$ in the reflected wave. A comparison of Figs. 9(a), (b), (c), and (d), along with consideration of frequency and transmission distance, reveals that the evolution pattern of OAM waves with $l=-2$ mirrors that of $l=-1$. Upon observing Figs. 9(e), (f), (g), and (h) and comparing them with Figs. 8(e), (f), (g), and (h), it was discovered that the reflection phase shifted from a single spin arm at $l=-1$ to two spin arms at $l=-2$.

When the MS topological is set to $l=+3$ and $RCP$ is utilized for vertical incidence, the results are presented as follows:
\begin{figure}[ht]
\centering{
\includegraphics[width=\linewidth]{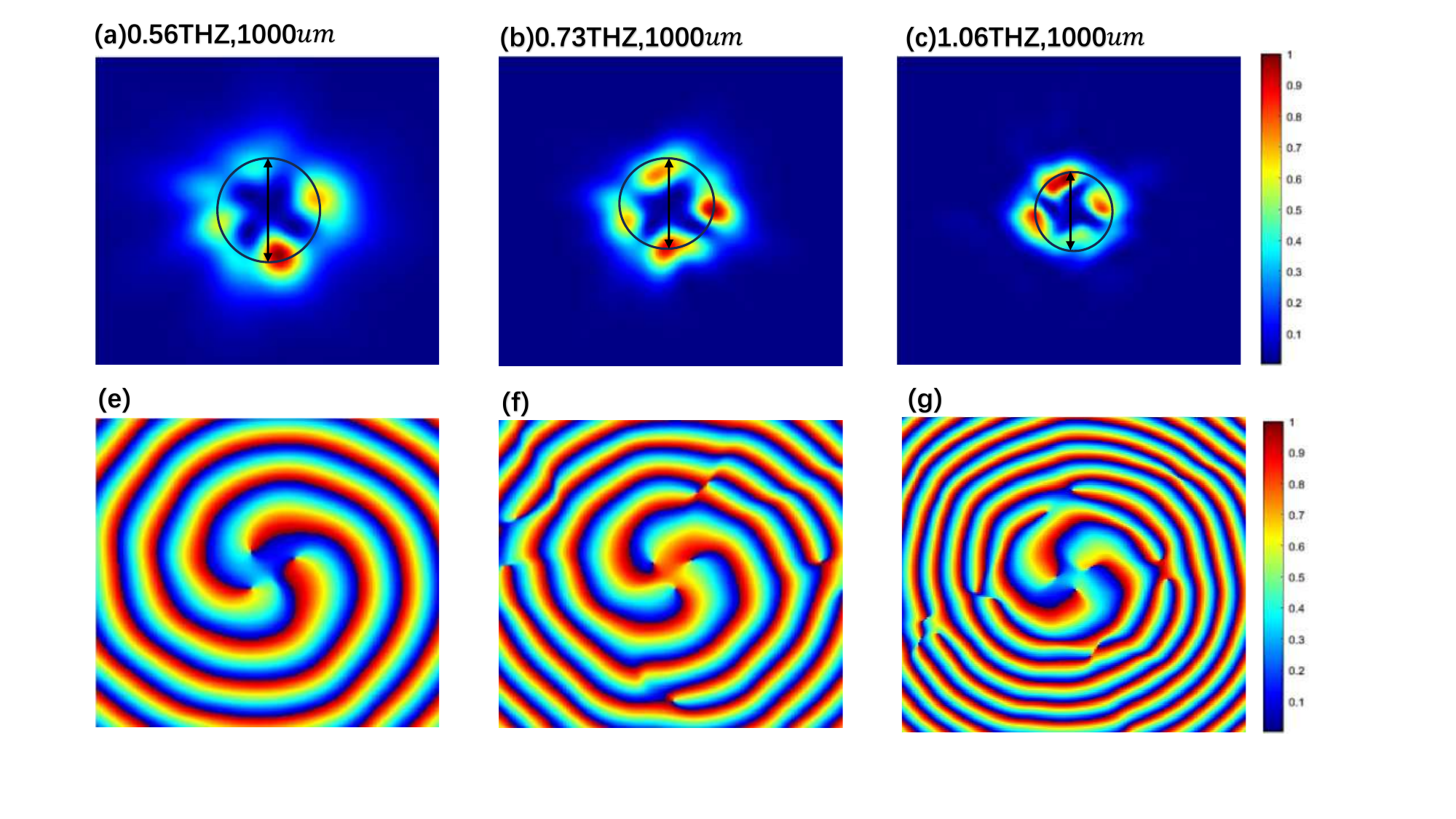}}
\caption{ The normalized reflection amplitudes  for $l=-3$ at $L=1000\mu m$ are presented for three different frequencies:(a)$f=0.56THz$, (b)$f=0.73THz$, (c)$f=1.06THz$. The corresponding reflection phases for (a), (b), (c) are shown in (d), (e), (f), respectively.}
\label{fig:FIG.10.}
\end{figure}

The results presented above were obtained when incoming $RCP$ Gaussian wave was used, with the outgoing wave also being $RCP$ and carrying a phase of $e^{-i3\psi}$. 

In addition, we present the simulation results for $LCP$ incident, as depicted in Fig.11. It is evident that the number of spin arms increases with the number of topological charges, with the increasing of topological charge (absolute value), the broadening of vortex wave is more obvious. When $LCP$ and $RCP$ incident upon the $l=+1$ order MS respectively, the twist direction of the phase spiral arms is opposite, and the topological charges of the generated vortex waves are equal in magnitude but opposite in sign. This is in agreement with the anticipated outcomes. In addition, the annular light intensity distribution appears obvious distortion with the increasing of topological charge, which is caused by the generation of phase noise due to the insufficient resolution of OAM phase structure.

\begin{figure}[ht]
\centering{
\includegraphics[width=\linewidth,height=0.5\linewidth]{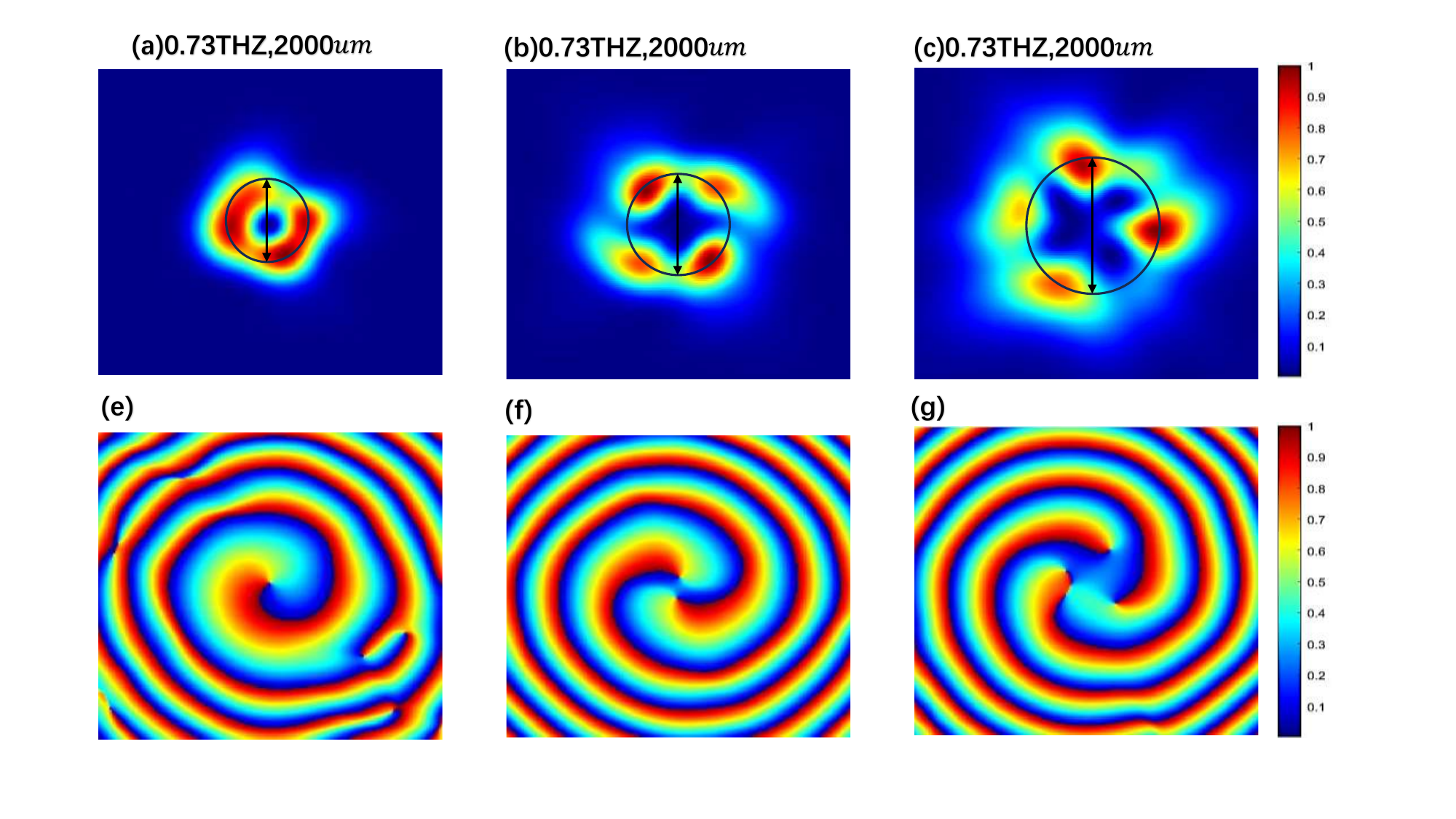}}
\caption{ The normalized reflection amplitudes at $L=2000\mu m$ are presented for three different topological charges:(a)$l=+1$,(b)$l=+2$, (c) $l=+3$.The corresponding reflection phases for (a), (b), (c) are shown in (d), (e), (f), respectively.}
\label{fig:FIG.11.}
\end{figure}

According to the simulation results, the reflective MS designed for $l=+1$ ,$l=+2$ and $l=+3$ can generate vortex wave in the frequency range of $0.52THz-1.1THz$.

\section{Mode Purity}
To assess the mode purity of vortex waves possessing various topological charges, we utilize the helical harmonic function $e^{il\varphi}$ to represent the complex amplitude of the wave.

\begin{align}
\begin{split}
E(x,y,z)=\frac{1}{\sqrt{2\pi} } \sum_{l=-\infty }^{+\infty } a_{l} (r,z)e^{il\varphi } 
\end{split}
\end{align}
the expansion coefficient is:
\begin{align}
\begin{split}
a_{l} (r,z)=\frac{1}{\sqrt{2\pi} }\int_{0}^{2\pi} E(x,y,z)e^{-il\varphi } d\varphi
\end{split}
\end{align}
From this, the energy of helical harmonic can be obtained:
\begin{align}
\begin{split}
 W_{l}=\int_{0}^{\infty } \left | a_{l} (r,z)^2 \right | rdr
\end{split}
\end{align}

\begin{align}
\begin{split}
P_{l}=\frac{W_{l}}{\sum_{q=-\infty }^{+\infty }W_{q}}
\end{split}
\end{align}
The orbit angular momentum spectrum can be obtained. 

In this paper, we tested the purity of MS($+1,+2,+3$) by exposing it to frequencies of $0.56THz$ and $1.06THz$, under the incidence of $RCP$ and $LCP$, respectively.
\begin{figure}[ht]
\centering{
\includegraphics[width=\linewidth,height=.6\linewidth]{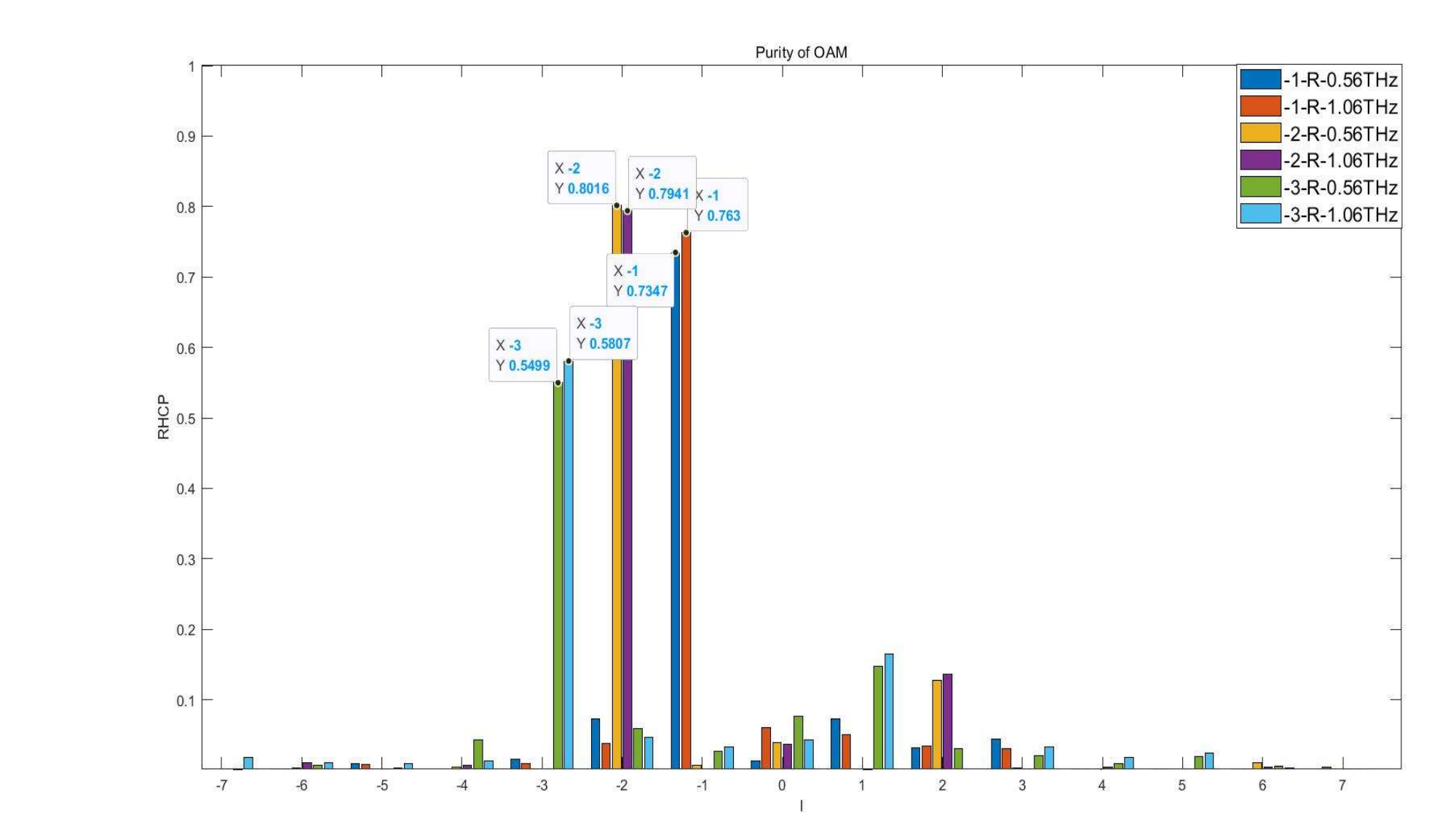}\put(-200,120){(a)}
\label{fig:FIG.12.a}
\includegraphics[width=\linewidth,height=.6\linewidth]{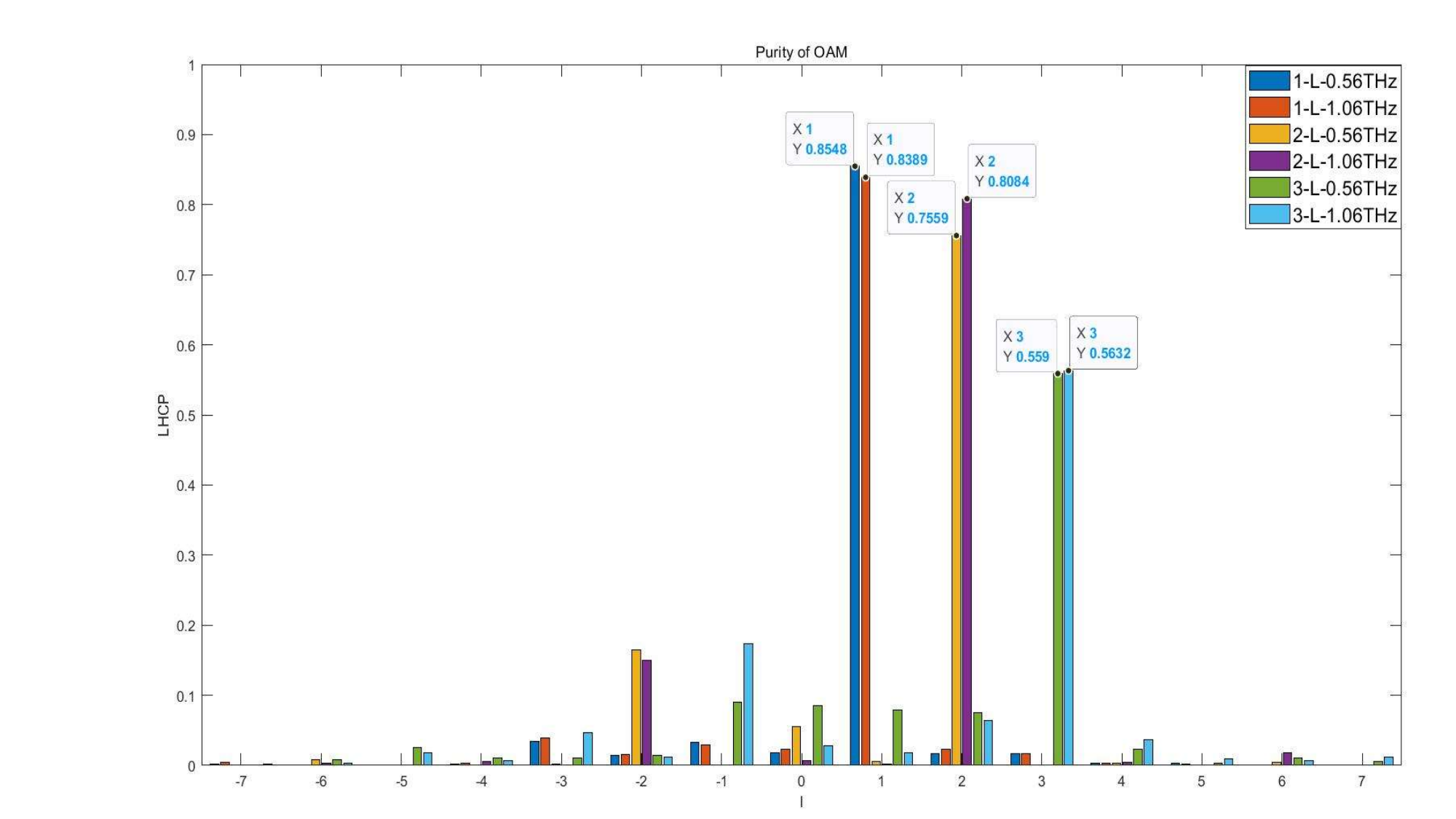}\put(-200,120){(b)}
\label{fig:FIG.12.b}}
\caption{Purity of OAM mode for: (a) $RCP$ and (b) $LCP$ with incident topological charges of $\pm1,\pm2,\pm3$  MS at frequencies of $0.56THz$ and $1.06THz$, respectively.}
\end{figure}

As shown in Fig. 12, the legend "$-1-R-0.56THz$" indicates that the $RCP$ carries a topological charge of $-1$ at $0.56THz$, and other legends are interpreted similarly. 
We can observe that the phase of the vortex wave changes when $RCP$ and $LCP$ are incident, respectively. Specifically, in the case of incident $RCP$, the outgoing $RCP$ carries topological charges of $l=-1,-2,-3$, and its purity at the frequencies of $0.56THz$ and $1.06THz$ is as follows: $73.47\%$, $76.3\%$, $80.16\%$, $79.41\%$, $54.99\%$, $58.07\%$. On the other hand, in the case of incident $LCP$, the outgoing $LCP$ carries topological charges of $l=+1,+2,+3$, and its purity at the frequencies of $0.56THz$ and $1.06THz$ is as follows: $85.48\%$, $83.89\%$, $75.59\%$, $80.84\%$, $55.9\%$, $56.32\%$.

The increase in topological charge and quantization loss leads to the generation of phase noise, which distorts the distribution of wave intensity and subsequently reduces the purity.

\section{Comparative Table for the Performance Evaluation of MS Vortex Generators}
To assess the performance of the method (MS) introduced in this paper, a comparison table with previous studies is provided.
\begin{table}[t]
\centering
\caption{Terahertz band reflective  geometric  phase MS vortex generators and their associated key parameters.} 
\label{samples}
\vspace{5pt}
\begin{tabular}{cccc}
\toprule[2pt] 
Frequency(THz)  &$l$ &Purity(\%)  \\
\midrule[1pt]
0.3-0.45  &$\pm1,\pm2$ &67-90.5 \cite{41} \\
0.5-1.06  &$\pm1,\pm2,\pm3$ &54.99-85.48(our work) & \\
0.45,0.75  &$\pm1,\pm2$  &66-94  \cite{51} \\
0.706,1.143,1.82  &$\pm1,\pm2,\pm3$  &60-85.1  \cite{50} \\

\bottomrule[2pt]
\end{tabular}
\end{table}

Our work encompasses a broader scope and is more continuous in comparison to previous efforts, while also boasting a simplified structure. This allows for cost-effective and more accessible design of MS devices.

\section{Conclusion}
In summary, we have introduced a novel terahertz broadband vortex generator leveraging geometric phase MS. This MS is capable of generating high-quality vortex waves within the frequency range of $0.56THz$ to $1.1THz$. The MS unit cell is primarily composed of a DSRR, a medium in the middle, and a metal substrate. It effectively converts $x(y)$-linear polarization into the corresponding cross-polarization. When circularly polarized (CP) waves interact with unit cells containing variously oriented DSRRs, the normalized amplitude of the reflected co-polarized CP remains above $85\%$. By rotating the DSRR, the reflection amplitude satisfies the coverage of a $2\pi$ phase. Additionally, we elaborate on the resonance mechanism of the unit cell.
Subsequently, we conducted numerical simulations for MS with topological charges of $+1,+2,+3$ within the corresponding frequency range. Through the incidence of $RCP$ and $LCP$ waves, respectively, we successfully generated $RCP$ vortex waves with topological charges of $+1,+2,+3$ and $LCP$ vortex waves with topological charges of $+1,+2,+3$. Notably, within the frequency range of interest, the mode purity of vortex waves with different topological charges exceeds $54\%$.

The unit cell proposed in this work boasts a straightforward design, wide frequency coverage, and high reflection amplitude. Furthermore, during the parameter optimization process, the structural parameter changes and numerical simulation results exhibit remarkable fault tolerance, facilitating its preparation. Based on its simple structure, the MS can be designed with multiple resonance outcomes to achieve multi-broadband operation, which warrants further investigation.

\section*{Acknowledgments}Financial support from the project funded by the State Key Laboratory of Quantum Optics and Quantum Optics Devices, Shanxi University, Shanxi, China (Grants No. KF202004 and No. KF202205).

\bibliography{main}
\end{document}